\documentclass{article}

\usepackage{PRIMEarxiv}

\usepackage{xcolor} 
\usepackage[utf8]{inputenc} 
\usepackage[T1]{fontenc}    
\usepackage{hyperref}       
\usepackage{url}            
\usepackage{booktabs}       
\usepackage{amsmath}
\usepackage{amsfonts}       
\usepackage{nicefrac}       
\usepackage{microtype}      
\usepackage{lipsum}
\usepackage{fancyhdr}       
\usepackage{graphicx}       
\usepackage[ruled,vlined]{algorithm2e}
\SetAlgoNoLine
\graphicspath{{media/}}     

\title{Keyed Chaotic Dynamics for Privacy-Preserving Neural Inference}

\author{
  Peter David Fagan \\
  School of Informatics\\
  University of Edinburgh\\
  United Kingdom \\
  \texttt{p.d.fagan@ed.ac.uk} \\
}

\begin{document}
\maketitle

\begin{abstract}
Neural network inference typically operates on raw input data, increasing the risk of exposure during preprocessing and inference. Moreover, neural architectures lack efficient built-in mechanisms for directly authenticating input data. This work introduces a novel encryption method for ensuring the security of neural inference. By constructing key-conditioned chaotic graph dynamical systems, we enable the encryption and decryption of real-valued tensors within the neural architecture. The proposed dynamical systems are particularly suited to encryption due to their sensitivity to initial conditions and their capacity to produce complex, key-dependent nonlinear transformations from compact rules. This work establishes a paradigm for securing neural inference and opens new avenues for research on the application of graph dynamical systems in neural network security.
\end{abstract}

\section{Introduction}
Modern machine learning systems increasingly operate in environments where user privacy, data security, and attribution are critical. Inference pipelines often transmit sensitive user data to hosted models; in these contexts, users must trust that their data are handled securely, and that their contributions are not exposed, misused, or misattributed. Existing approaches to privacy-preserving inference often rely on heavyweight cryptographic methods such as homomorphic encryption \cite{gilad2016cryptonets}, and secure multi-party computation \cite{mohassel2017secureml} which require significant computational overhead or infrastructure changes, limiting their practicality in real-world systems and motivating the need for lightweight, deployable alternatives.

To address this gap, we seek a practical solution for securing real-valued tensors during inference—one that supports both data privacy and attribution without disrupting existing model architectures. Ideally, such a solution would act as a cryptographic wrapper for tensors, applied either prior or post inference, without incurring significant computational overhead.

This work introduces a framework for constructing chaotic graph dynamical systems and applying them to encrypt and decrypt tensors used in machine learning pipelines, enabling privacy-preserving and traceable manipulation of data representations. Each dynamical system is constructed from a cryptographic key and is represented as a computation graph that can readily be incorporated into existing neural architectures. 

The main contributions of this work are:

\begin{itemize}
    \item A framework for constructing chaotic graph dynamical systems from cryptographic keys. 
    \item A demonstration of the application of chaotic graph dynamical systems for encrypting and decrypting tensors with a focus on inference pipelines.
    \item A discussion of the implications of this paradigm for future research directions in privacy preserving machine learning.
\end{itemize}

This framework bridges ideas from cryptography and dynamical systems offering a new lens on how models and data can interact securely. Unlike traditional cryptographic methods, this approach is architecture-agnostic, low-overhead, and enables per-user control over inference contributions. The following sections outline the approach, initial validation experiments and directions for future work. 

\section{Background and Related Work}

\subsection{Cryptography}

Modern cryptographic systems rely on key-dependent, invertible transformations that enable secure and authorized access to data. These transformations, when combined with additional primitives such as cryptographic hashes and digital signatures, support confidentiality, integrity, and authenticity across a wide range of secure communication protocols \cite{stallings2017cryptography}. While most cryptographic methods operate on discrete data (e.g., byte strings), the key ideas — particularly keyed deterministic transforms and collision-resistant signatures — are relevant when designing mechanisms for secure computation in neural systems.

Two foundational building blocks in traditional cryptography are symmetric-key encryption and cryptographic hash functions. Symmetric schemes such as AES and HMAC rely on a shared secret key to encrypt or authenticate data, ensuring that only entities with access to the key can reproduce or verify the transformation \cite{nist-aes, nist-hmac}. These methods are computationally efficient and deterministic, but are designed for bitwise or block-level data and require quantization or serialization to operate on high-dimensional tensors. While invertible, they are not natively compatible with neural network pipelines, which often operate on continuous, real-valued representations during inference.

Cryptographic hash functions such as SHA-256 offer another relevant abstraction: they produce outputs that are deterministic, sensitive to small input changes, and hard to invert or collide \cite{nist-sha}. However, they map inputs to fixed-length binary strings and do not support inverse gradients, interpolation, or alignment with downstream learning tasks.

The method introduced in this work draws inspiration from these primitives but is fundamentally distinct: it operates in continuous space and supports tensor-valued inputs. By using keys to construct graph dynamical systems that encrypt tensors, the method introduces cryptographically inspired properties, such as key-dependent transformation and input sensitivity, while remaining lightweight and compatible with standard model architectures.

\subsection{Privacy-Preserving Inference}
As machine learning systems become increasingly deployed in real-world, cloud-based, and client-server settings, privacy-preserving inference has emerged as a critical area of research. In many applications, users transmit sensitive data, such as biometrics, medical signals, personal text, or proprietary feature representations, to centralized models for inference. This has raised concerns not only about data exposure, but also about the reusability, traceability, and control of these contributions once they enter the inference pipeline.

A number of approaches have been proposed to mitigate these risks, but each carries substantial trade-offs in performance, infrastructure complexity, or applicability. Techniques such as homomorphic encryption (HE) \cite{gilad2016cryptonets} allow models to perform inference directly on encrypted data, preserving confidentiality without revealing inputs to the model host. However, HE-based approaches are often computationally intensive and have orders of magnitude slower model inference, limiting their deployment to narrow use cases. Similarly, secure multi-party computation (SMPC) distributes computation across parties such that no individual party has full access to the data, but these methods typically require multiple rounds of communication, synchronized computation, and significant coordination overhead \cite{mohassel2017secureml}.

Other approaches focus on statistical privacy. For example, differential privacy (DP) introduces controlled noise into model training or inference pipelines to mask individual contributions \cite{abadi2016deep}. While DP offers formal privacy guarantees, it is stochastic, lossy, and often difficult to apply meaningfully in one-off inference scenarios. Lastly, training-time watermarking techniques embed identifiable signatures into the model’s parameters during training, enabling ownership verification or post hoc misuse detection \cite{uchida2017embedding}. However, these methods are typically static, global, and tied to the model itself—making them difficult to extend to per-user or inference-time watermarking, and often requiring access to the model weights.

The approach introduced in this work departs from these paradigms. Rather than encrypting the model or introducing noise, it applies deterministic, key-conditioned transformations to tensors prior or post inference. These transformations are derived from chaotic dynamical systems, producing outputs that are key-specific, structurally transformed, and task-compatible (i.e., preserving shape and structure for downstream model compatibility). While the masking operation is algebraically reversible by the key holder (e.g. via replaying forward dynamics and performing subtraction), it requires a key as with traditional encryption methods like AES. As a result, the overall transformation remains opaque to unauthorized parties. The chaotic nature of the transformation ensures exponential sensitivity to small changes in initial state or key, rendering inversion impractical without full system knowledge—even though the transformation is technically deterministic. This enables lightweight, key-authenticated inference, where only correctly transformed inputs yield meaningful results, and offers tensor-level control over privacy, traceability, and secure data contribution—without requiring architectural changes or incurring the computational costs of conventional cryptographic methods.

\subsection{Chaotic Dynamical System}
Chaotic dynamical systems are processes that exhibit sensitive dependence on initial conditions, nonlinear evolution, and long-term unpredictability despite often being fully governed by known rules. They are widely studied in physics, biology, and nonlinear systems theory \cite{strogatz2024nonlinear}, and have seen limited but notable use in machine learning—primarily as components in reservoir computing.

In the context of this work, chaos is used as a security primitive with analogies drawn between the properties of chaotic dynamics and the properties of traditional encryption algorithms. Chaos in this work is identified by the presence of a positive Lyapunov exponent \cite{ott2002chaos}, indicating exponential sensitivity to small changes in initial conditions. The key properties of chaotic systems that make them suitable as a security primitive include:

\begin{itemize}
    \item Determinism: A fixed chaotic system produces consistent trajectories from identical inputs.
    \item Sensitivity to initial conditions: Small changes in initial state result in vastly different trajectories.
    \item Structural complexity: Chaotic transformations encrypt input tensors into seemingly unstructured forms, making them difficult to interpret or invert without full system knowledge.
\end{itemize}	

In traditional encryption algorithms such as AES, security is achieved through a series of key-dependent, invertible transformations—typically involving substitution-permutation networks and multiple rounds of mixing operations. While structurally different, the chaotic transformations used in this work serve a comparable role: they are deterministic, key-conditioned, and designed to obscure the relationship between input and output tensors. Like AES, small changes to the key or input yield drastically different outputs, ensuring unpredictability and resistance to unauthorized inversion. However, unlike AES, our transformations operate in continuous space, are differentiable, and integrate seamlessly into existing machine learning pipelines—enabling lightweight, task-compatible obfuscation without requiring model modification or cryptographic decoding layers.

\section{Keyed Chaotic Dynamics}

\begin{figure}[ht!]
    \centering
    \includegraphics[width=0.7\linewidth]{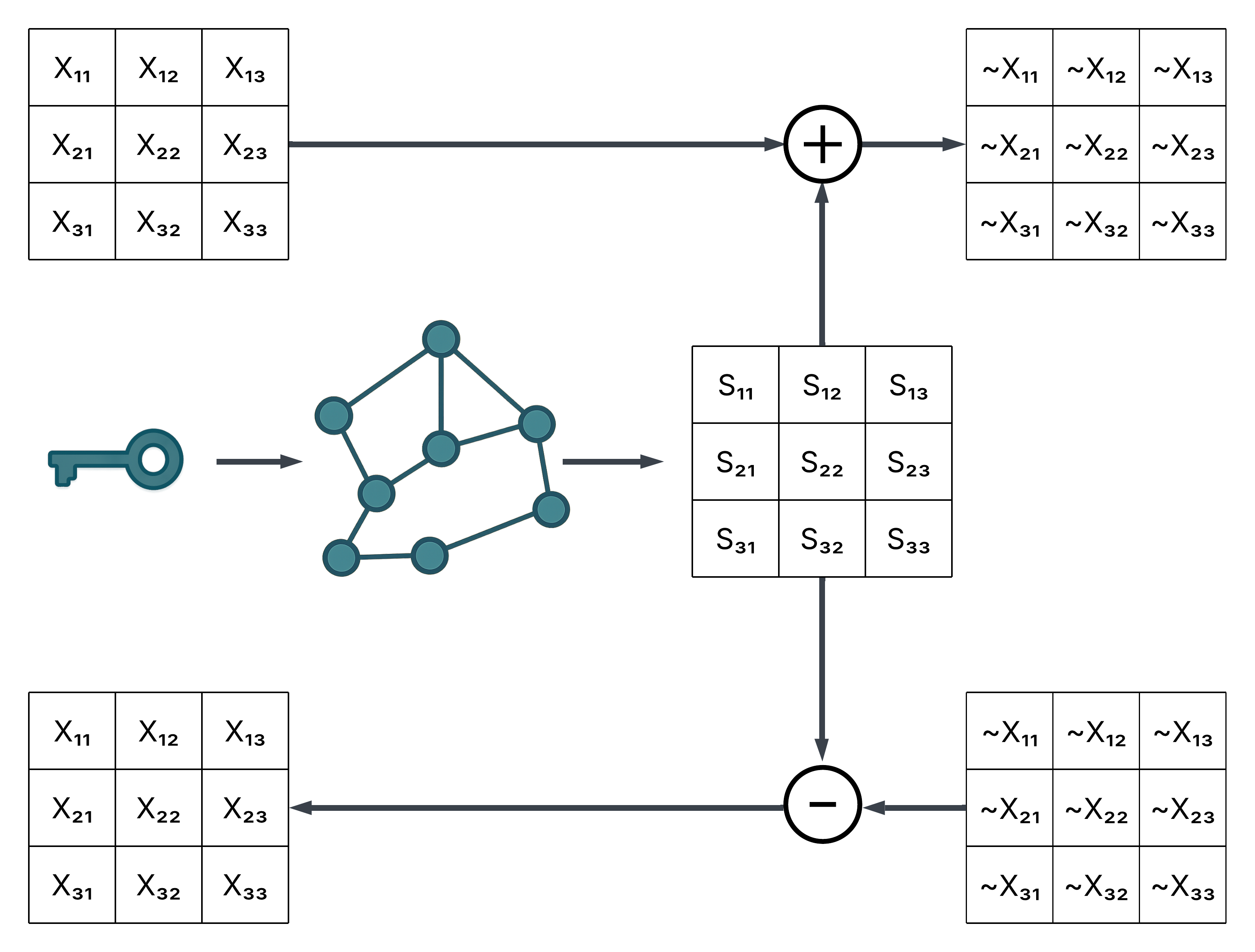}
    \caption{Overview of the keyed chaotic masking and unmasking process. A cryptographic key is used to instantiate a chaotic graph dynamical system, which produces a deterministic mask matrix $S$. During encryption (top path), this mask is element-wise added to the input tensor X, producing a masked tensor $\tilde{X} = X + S$. During decryption (bottom path), the same mask is regenerated using the shared key and subtracted from $\tilde{X}$ to recover the original input: $X = \tilde{X} - S$. The central graph diagram represents the chaotic system generated from the key, and the matrices represent tensor-valued data.}
    \label{fig:enter-label}
\end{figure}

\subsection{Key Hashing and Seed Derivation}
The chaotic graph dynamical system used in this work is deterministically instantiated from a cryptographic key through a structured key schedule and parameter sampling process. We assume the presence of a secure key exchange protocol, such as Elliptic Curve Diffie-Hellman (ECDH), to derive a shared secret key $k$. To prevent reuse of masks across different inputs or sessions, we also introduce a public nonce \( n \), which is combined with the key during subkey derivation. To securely and modularly configure the chaotic system from this key, we adopt a domain-specific key schedule inspired by symmetric encryption algorithms such as AES. We start by deriving multiple subkeys from the master key k by hashing together fixed domain labels and the nonce with the key:

\begin{equation}
\begin{aligned}
k_{\text{graph}}  &= \text{SHA-256}(\texttt{"graph"} \,\|\, k  \,\|\, n) \\
k_{\text{params}} &= \text{SHA-256}(\texttt{"params"} \,\|\, k \,\|\, n) \\
k_{\text{init}}   &= \text{SHA-256}(\texttt{"init"} \,\|\, k \,\|\, n)
\end{aligned}
\end{equation}

Each resulting subkey is then used to seed its corresponding sampling procedure: graph topology construction, chaotic map parameter selection, and initial condition generation. This separation improves modularity, ensures reproducibility, and avoids parameter entanglement between system components.

\subsection{Constructing Chaotic Graph Dynamical Systems}

The goal of this section is to construct a graph dynamical system whose behavior is key-conditioned and exhibits characteristics of chaos. This system serves as the foundation for a transformation mechanism that operates on input or output tensors, enabling secure and traceable inference without modifying neural network architectures.

\subsubsection{Graph Sampling}

The chaotic system used for transformation is implemented over a graph, where nodes represent internal state variables and edges determine the coupling between node dynamics. The topology of this graph is sampled deterministically from a cryptographic key, producing a fixed but key-specific structure that supports rich dynamical behavior when combined with nonlinear update rules.

Given a cryptographically derived subkey $k_{\text{graph}}$, obtained via SHA-256 of a labeled key expansion step, we treat it as a high-entropy seed for initializing a pseudorandom number generator (PRNG). This ensures deterministic graph sampling while maintaining strong key-dependent variation. If needed for compatibility with legacy APIs, a 32-bit integer seed can be extracted from the first four bytes of $k_{\text{graph}}$. The resulting PRNG instance is then used to construct the system’s graph topology \( G = (V, E) \), where \( |V| = d \) matches the dimensionality of the tensor being transformed. In practice we represent the graph with a single matrix $W \in \mathbb{R}^{d \times d}$ representing the coupling strengths between nodes.

In this work, adjacency matrices $A$ representing graph topologies are first deterministically sampled from classical random graph families, including Erdős–Rényi (ER) and Watts–Strogatz (WS) small-world models. For ER graphs, edges are sampled independently with fixed probability $p$, while small-world graphs are generated using a $k$-regular lattice with random rewiring governed by parameter $\beta$.

Having sampled an adjacency matrix $A$, edge weights which define coupling strengths between nodes are sampled. In the simplest case, a weight matrix $W$ is drawn from a uniform distribution over a fixed interval and subsequently the Hadamard product is taken between this matrix and the adjacency matrix as follows:

\begin{equation}
W \leftarrow W \odot A
\end{equation}

As with the sampling of adjacency matrices, the current framework generalizes to arbitrary weight sampling distributions, enabling finer control over spectral properties and interaction strength. The resulting key-specific graph serves as the backbone of the chaotic system, defining how local dynamics interact and evolve over time.

\subsubsection{Node Dynamics}
Each node in the system evolves according to a discrete-time update rule that combines a local nonlinear map with global coupling from neighboring nodes. The resulting dynamical behavior emerges from the interaction between the node-level nonlinearity and the graph structure. 

Let \( \mathbf{x}(t) \in \mathbb{R}^d \) represent the state of the system at time \( t \), and let \( W \in \mathbb{R}^{d \times d} \) denote the sampled weight matrix representing the graph. The state update proceeds according to:

\begin{equation}
\tilde{\mathbf{x}}(t+1) = f( \mathbf{x}(t), W; \phi) = f_{\text{local}}(\mathbf{x}(t); \phi) +  f_{\text{coupled}}(\mathbf{x}(t), W; \phi)
\end{equation}

where \( f \) denotes the chosen node-level update dynamics, and \( \phi \) represents dynamics-specific parameters. In most cases, \( f \) incorporates both local nonlinear updates and graph-mediated coupling via the weighted matrix.

In this work, several candidates for local nonlinear updates are used, including:

\begin{table}[h]
\centering
\renewcommand{\arraystretch}{1.4}
\begin{tabular}{|l|l|l|}
\hline
\textbf{Map Name} & \textbf{Update Rule} & \textbf{Output State Parameter} \\
\hline
\textbf{Logistic map} & 
$ x_{i+1} = r x_i (1 - x_i) $ & 
$ x_i \in (0, 1) $ \\
\hline
\textbf{Tent map} & 
$ x_{i+1} = 
\begin{cases} 
\frac{x_i}{\mu} & x_i < \mu \\
\frac{1 - x_i}{1 - \mu} & x_i \geq \mu
\end{cases} $ & 
$ x_i \in (0, 1) $ \\
\hline
\textbf{Baker's map} & 
$ 
(x_{i+1}, y_{i+1}) = 
\begin{cases} 
(s x_i, y_i) & x_i < s \\
((1 - s)x_i + s, (1 - s)y_i + s) & x_i \geq s 
\end{cases}
$ & 
$ x_i \in [0, 1] $ \\
\hline
\textbf{Standard map} & 
$\begin{aligned}
p_{i+1} &= p_i + K \sin(\theta_i) \mod 2\pi \\
\theta_{i+1} &= \theta_i + p_{i+1} \mod 2\pi
\end{aligned}$ & 
$ \theta_i \in \mathbb{R} $ \\
\hline
\textbf{Arnold's cat map} & 
$ 
\begin{bmatrix}
x_{i+1} \\
y_{i+1}
\end{bmatrix}
=
\begin{bmatrix}
1 & 1 \\
1 & 2
\end{bmatrix}
\begin{bmatrix}
x_i \\
y_i
\end{bmatrix}
\mod 1
$ & 
$ x_i \in [0, 1] $ \\
\hline
\end{tabular}
\caption{Summary of candidate node-level chaotic maps used in this work.}
\end{table}

Different methods of coupling between nodes can be applied to each of the local nonlinear update rules. In this work we choose diffusive coupling between nodes which is defined as: 

\begin{equation}
\tilde{x}_i(t+1) = f_{\text{local}}(x_i(t); \phi) + \sum_{j=1}^d W_{ij} \left( f_{\text{local}}(x_j(t); \phi) - f_{\text{local}}(x_i(t); \phi) \right)
\end{equation}

At runtime, the update rule is selected from a library of registered dynamics, and its parameters are sampled from ranges known to result in bounded chaotic dynamics. This selection is deterministically conditioned on the cryptographic key, enabling the transformation system to be both modular and reproducible across users and devices.

\subsubsection{Chaotic Dynamics Diagnostics}
To ensure that the graph dynamical systems constructed using cryptographic keys exhibit chaotic behavior, we perform a series of diagnostics drawn from nonlinear systems theory. Chaos is typically defined by the presence of a positive Lyapunov exponent, indicating exponential sensitivity to small differences in initial conditions. This property is essential for ensuring that similar inputs or neighboring keys result in divergent transformation trajectories an important property for ensuring the security of this encrpytion method.

To assess whether the system exhibits chaotic behavior, we estimate the largest Lyapunov exponent using a standard two-trajectory method. We initialize two states: \( \mathbf{x}_0 \) and a perturbed version \( \hat{\mathbf{x}}_0 = \mathbf{x}_0 + \epsilon \), where \( \epsilon \) is a small positive scalar. Both trajectories are evolved synchronously under the same dynamical system. At each step, the separation \( \delta(t) = \hat{\mathbf{x}}(t) - \mathbf{x}(t) \) is measured, and the Lyapunov exponent is estimated by averaging the log growth rate:
\[
\lambda \approx \frac{1}{T} \sum_{t=1}^T \log \left( \frac{ \| \delta(t) \| }{ \epsilon } \right)
\]
To ensure numerical stability, the perturbed state is renormalized after each step:
\[
\hat{\mathbf{x}}(t) \leftarrow \mathbf{x}(t) + \epsilon \cdot \frac{ \delta(t) }{ \|\delta(t)\| + \varepsilon }
\]
where \( \varepsilon \) is a small constant to avoid division by zero. A consistently positive value of \( \lambda \) indicates exponential divergence and confirms the presence of chaotic dynamics.

\subsection{Encryption Algorithm}

This section consolidates the components described above into a complete encryption and decryption procedure. Given a cryptographic key, a deterministic chaotic graph dynamical system is instantiated by deriving subkeys and sampling graph structure and node dynamics. Forward dynamics is then simulated to produce a structured mask that is applied to the input tensor via element-wise addition in the case of encryption and subtraction in the case of decryption. The resulting process is lightweight, reproducible, and key-dependent—offering functional privacy, it can also be incorporated directly into existing neural architectures as a layer without requiring retraining of the model architecture. The full algorithm is outlined below:

\begin{algorithm}[h]
\KwIn{Input tensor \( X \), secret key \( k \), public nonce \( n \)}
\KwOut{Masked tensor \( \tilde{X} \) or recovered tensor \( X \)}

Derive subkeys \( k_{\text{graph}}, k_{\text{params}}, k_{\text{init}}, k_{\text{noise}} \) using SHA-256:\;
\ForEach{label \( i \in \{\texttt{graph}, \texttt{params}, \texttt{init}, \texttt{noise} \} \)}{
    \( k_i \leftarrow \text{SHA-256}(\texttt{label}_i \,\|\, k \,\|\, n) \)\;
}

Construct chaotic graph dynamical system:\;
\Indp
Sample graph topology \( G \) from \( k_{\text{graph}} \)\;
Configure parameters and initialize state from \( k_{\text{params}}, k_{\text{init}} \)\;
Seed a pseudorandom number generator with \( k_{\text{noise}} \) to generate deterministic noise\;
\For{each timestep \( t = 1 \) to \( T \)}{
    Apply node updates using local dynamics and diffusive coupling\;
    Inject deterministic noise into each node state at timestep \( t \)\;
}
Aggregate final node states into mask matrix \( S \in \mathbb{R}^{n \times d} \)\;
\Indm
\BlankLine
\textbf{Encryption:} \( \tilde{X} \leftarrow X + S \)\;
\textbf{Decryption:} \( X \leftarrow \tilde{X} - S \)\;

\caption{Keyed Chaotic Dynamics (Encryption and Decryption with Deterministic Noise)}
\end{algorithm}






\section{Applications}

This section outlines practical use cases enabled by the proposed transformation framework. While experimental validation is ongoing, the design properties of the system—key-conditioned masking, deterministic reversibility, and compatibility with real-valued tensor representations—make it suitable for deployment in a range of privacy-sensitive machine learning settings. In all cases, we assume a shared secret key \( k \) is securely negotiated between parties (e.g., via ECDH) and that a public nonce \( n \) is used to ensure per-input uniqueness.

\subsection{Key-Authenticated Neural Network Inference}

In cloud-hosted or API-based inference systems, it is often desirable to restrict access to only authorized users—both for security and for usage control. Traditional access control mechanisms typically operate at the network or user level—using API keys, authentication tokens, or IP-based restrictions. While these methods can identify the requester, they provide no cryptographic binding between the user and the data submitted to the model. As a result, once access is granted, any input can be submitted, including replayed, forged, or malicious content.

The proposed transformation scheme shifts authentication to the data level by requiring that inputs be correctly transformed using a user-specific, key-conditioned chaotic system. This approach ensures that only users with the correct cryptographic key can produce inputs that yield valid model outputs, effectively binding model access to both the identity and integrity of the data itself. It offers an additional layer of protection against misuse, credential leakage, and unauthorized inference—without requiring changes to the model or encrypted communication channels.

Let \( \mathbf{X} \in \mathbb{R}^{n \times d} \) be a sequence of token embeddings. A user constructs a chaotic graph dynamical system from a cryptographic key \( k \) and nonce $n$, and generates a deterministic mask \( \mathbf{S}_k \in \mathbb{R}^{n \times d} \). The input is encrypted client-side via:
\[
\tilde{\mathbf{X}} = \mathbf{X} + \mathbf{S}_k
\]
and transmitted to the model. At inference time, the model internally reconstructs the true input by regenerating the same chaotic system and subtracting the mask:
\[
\mathbf{X} = \tilde{\mathbf{X}} - \mathbf{S}_k
\]
Only if the correct key is used will this inversion yield a valid input. Any incorrectly masked input results in faulty or degraded inference, effectively binding model access to possession of the correct key.

Optionally, the model can also encrypt its outputs using a second chaotic trajectory:
\[
\tilde{\mathbf{y}} = \mathbf{y} + \mathbf{S}'_k
\]
The client then performs the inverse:
\[
\mathbf{y} = \tilde{\mathbf{y}} - \mathbf{S}'_k
\]
This creates a fully authenticated inference pipeline, where both input and output representations are cryptographically tied to the user’s key. This transformation operates independently of network-layer encryption, and can be combined with standard secure transport protocols to protect both data in transit and storage. 

\subsection{Privacy-Preserving Data Sharing and Inference}

In both inference and collaborative learning settings, real-valued embeddings may contain sensitive information (e.g., personal text, biometrics, or proprietary features). The proposed transformation enables users to encrypt these representations locally using a key-conditioned chaotic system before transmitting them to a server or aggregator.

The encrypted input \( \tilde{X} = X + S_k \) retains the structure and shape required for downstream tasks. Because the transformation is deterministic and reversible with the key and nonce, the server can reconstruct the original input for inference or training, this is especially relevant in TEE environments where the computational graph representing the encryption/decryption algorithm is not easily interpreted.

This enables secure, per-user control over data contributions—offering lightweight protection against inspection, misuse, or unauthorized reuse.

\section{Security Considerations}

\subsection{Threat Model}

This framework is designed as a lightweight privacy mechanism for inference-time data obfuscation and input authentication. It aims to:
\begin{itemize}
    \item Prevent unauthorized users from generating valid inputs to a model without access to a cryptographic key.
    \item Obfuscate real-valued inputs during transit or storage without modifying model architecture.
    \item Enable reversible masking for authorized parties in possession of the key.
\end{itemize}

We assume a standard threat model where:
\begin{itemize}
    \item The shared key \( k \) is negotiated securely (e.g., via Elliptic Curve Diffie-Hellman).
    \item The attacker has access to the masked tensor \( \tilde{X} \), but not the key or the internal configuration of the mask-generating chaotic system.
    \item The system is not used in repeated or low-entropy input settings without additional safeguards.
\end{itemize}

\subsection{Security Properties}

The proposed transformation offers the following functional guarantees:
\begin{itemize}
    \item \textbf{Key- and nonce-conditioned obfuscation:} The transformation from \( X \) to \( \tilde{X} \) is deterministic and reversible only with the correct cryptographic key \( k \) and associated nonce \( n \). This ensures unique masks per input and prevents mask reuse attacks, even when the same key is reused across queries.

    \item \textbf{Input authentication:} Only users in possession of the correct key and nonce can produce valid masked inputs that result in successful model inference, effectively binding access to possession of the correct credentials.

    \item \textbf{Nonlinear unpredictability:} The chaotic update rule introduces exponential sensitivity to both key and input perturbations. The injection of deterministic, key-derived noise at each timestep further increases trajectory divergence, degrading the feasibility of surrogate inversion or pattern learning.

    \item \textbf{Model compatibility:} The transformation operates entirely in continuous space and is represented by a computation graph. It integrates seamlessly into existing deep learning pipelines and can be implemented efficiently on hardware accelerators such as GPUs using CUDA kernels or fused operations, enabling low-latency deployment even in real-time inference settings.
\end{itemize}

\subsubsection{Implementation Considerations for Secure Deployment}

The transformation can be efficiently implemented as a fused CUDA kernel to minimize memory access overhead and latency. However, this design choice introduces additional considerations: GPU memory and execution contexts are shared across processes, and on systems lacking strict memory isolation, adversaries with privileged access may attempt to inspect device memory or leverage side-channel leakage (e.g., timing or power-based signals) to infer key-conditioned behavior. While such attacks require elevated privileges and are not typically feasible in hosted model settings, developers deploying this method in untrusted or multi-tenant environments should consider GPU sandboxing, memory zeroing, and kernel isolation practices. Integrating the transformation within a trusted execution environment (TEE) or extending it with secure enclaving techniques may offer additional protection against kernel-level inspection.

\subsection{Limitations and Attack Vectors}

This transformation is best viewed as a functional privacy primitive for continuous tensor data—analogous in structure to symmetric encryption schemes such as AES, but adapted for real-valued, neural architectures. While it does not aim to replace formal cryptographic methods in settings requiring ciphertext indistinguishability, it offers a complementary security protocol for scenarios involving neural inference. Potential vulnerabilities include:

\begin{itemize}
    \item \textbf{Model inversion attacks (unlikely):} An adversary with full access to masked inputs and model outputs could theoretically attempt to learn a surrogate inverse mapping. However, the transformation is highly nonlinear, key-conditioned, and exhibits exponential sensitivity to perturbations via chaotic dynamics. Additionally, the injection of deterministic, per-timestep noise renders the resulting mask trajectories resistant to approximation, making such attacks computationally infeasible in practice.

    \item \textbf{Key compromise:} The transformation is fully reversible by design. If the cryptographic key and associated nonce are leaked or mishandled, the original input can be trivially recovered by an adversary. Standard key management practices (e.g., ephemeral session keys, ECDH negotiation) should be used to mitigate this risk.

    \item \textbf{Lack of formal obfuscation guarantees:} The transformation depending choice of dynamics applies a bijective, key-conditioned nonlinear mapping over continuous, real-valued tensors. While the resulting representations are semantically obfuscated and reversible only by authorized parties, the system does not rely on finite-field cryptographic primitives and does not satisfy formal definitions of indistinguishability. As such, it should be viewed as a structured, functional privacy mechanism rather than a substitute for conventional encryption in adversarial threat models.
\end{itemize}

\subsection{Integration with Secure Execution}

This system is intended as a complementary privacy layer. In high-risk deployments, it can be combined with:
\begin{itemize}
    \item Standard transport-layer encryption (e.g., TLS)
    \item Trusted Execution Environments 
    \item Ephemeral key derivation or key rotation schemes
    \item Obfuscation techniques such as dropout, quantization, or learned masking
\end{itemize}

\subsection{Future Directions}

Future work includes:
\begin{itemize}
    \item Formal analysis of the system’s resistance to statistical and adversarial attacks.
    \item Investigation of dynamic or learned chaotic systems for adaptive masking.
    \item Exploration of surrogate model defenses and integration into split-learning architectures.
\end{itemize}

This framework is best viewed as a flexible privacy primitive that enables data-level masking and authentication with minimal architectural assumptions. It is particularly useful in scenarios where full encryption is infeasible or unnecessary, but lightweight, key-bound transformations are desirable.

\section{Discussion}

This work introduces a new class of tensor-level transformations based on keyed chaotic dynamical systems, designed to apply key-conditioned, reversible obfuscation to real-valued tensors within modern machine learning pipelines. The system is lightweight, deterministic, and modular—requiring no changes to model architecture and operating natively in continuous space. It is amenable to efficient GPU execution, including fused CUDA kernel implementations for real-time inference masking.

The transformation provides functional security guarantees such as cryptographic key- and nonce-conditioned masking, semantic obfuscation, and trajectory-level unpredictability. It supports both input authentication and reversible transformation, enabling inference-time privacy and secure data contribution without disrupting model compatibility. While the system is reversible by design for authorized users, its behavior in adversarial settings remains to be evaluated. In particular, the method does not rely on formal encryption primitives over finite fields and may be vulnerable to misuse if keys are leaked or mask reuse occurs without appropriate nonce management. As such, the transformation is best viewed as a complementary privacy layer—positioned between full encryption and raw data sharing—and can be combined with techniques such as differential privacy, trusted execution environments, or encrypted transport protocols in high-assurance contexts.

The design of the chaotic system (e.g., graph topology, local map, coupling strategy) plays a critical role in shaping both the security and representational properties of the transformation. This work restricts itself to a curated set of update rules with empirically verified chaotic behavior, but future directions include optimizing or learning system configurations from data, developing differentiable chaos layers, or adversarially training transformation functions to withstand surrogate model attacks.

More broadly, this work suggests that chaotic dynamical systems—traditionally studied in the context of memory, control, or biological computation—can be reimagined as lightweight, cryptographically inspired building blocks for AI systems. As machine learning continues to interface with sensitive and user-controlled data, techniques that support secure signaling, data attribution, and behavioral integrity will be increasingly essential. This framework provides one such path forward, and opens several directions for continued exploration.

\section{Conclusion}

This work introduces a novel framework for applying keyed chaotic dynamical systems to tensor transformations in machine learning. By leveraging deterministic chaos seeded from cryptographic keys and public nonces, the method enables encryption/decryption and authenticated inference based on keys without requiring retraining or alterations to original model architectures. The transformation operates directly on real-valued tensors in continuous space, making it compatible with modern neural pipelines and efficient to implement on hardware accelerators.

The proposed system serves as a functional privacy mechanism: key- and nonce-conditioned, reversible, and structurally opaque to unauthorized users. It offers a lightweight, modular alternative to traditional cryptographic methods for use cases involving secure inference, controlled data contribution, and user-authenticated API access.

As machine learning systems increasingly interact with sensitive and user-owned data, new methods that enable secure, interpretable, and user-aligned processing will become essential. This framework contributes one such approach, grounded in dynamical systems theory and designed for integration into existing model infrastructure. Validation experiments and empirical evaluations are ongoing. This preprint is shared to disseminate the concept, gather community feedback, and guide future research into learned chaotic dynamics, adversarial robustness, and secure deployment practices.

\section{Acknowledgements}

The author gratefully acknowledges their UKRI PhD studentship at the University of Edinburgh, which supported prior research training and early exploration of related ideas. This work was completed independently during a period of academic leave. The author also acknowledges the use of large language models (LLMs) to support drafting, editing, and idea development during the research and writing process.

\bibliographystyle{unsrt}  
\bibliography{references}

\begin{thebibliography}{10}

\bibitem{gilad2016cryptonets}
Ran Gilad-Bachrach, Nathan Dowlin, Kim Laine, Kristin Lauter, Michael Naehrig, and John Wernsing.
\newblock Cryptonets: Applying neural networks to encrypted data with high throughput and accuracy.
\newblock In {\em Proceedings of the 33rd International Conference on Machine Learning (ICML)}, pages 201--210, 2016.

\bibitem{mohassel2017secureml}
Payman Mohassel and Yupeng Zhang.
\newblock Secureml: A system for scalable privacy-preserving machine learning.
\newblock In {\em IEEE Symposium on Security and Privacy (S\&P)}, pages 19--38, 2017.

\bibitem{stallings2017cryptography}
William Stallings.
\newblock {\em Cryptography and Network Security: Principles and Practice}.
\newblock Pearson, 7th edition, 2017.

\bibitem{nist-aes}
National~Institute of~Standards and Technology.
\newblock Fips pub 197: Advanced encryption standard (aes).
\newblock \url{https://nvlpubs.nist.gov/nistpubs/FIPS/NIST.FIPS.197.pdf}, 2001.

\bibitem{nist-hmac}
National~Institute of~Standards and Technology.
\newblock Fips pub 198-1: The keyed-hash message authentication code (hmac).
\newblock \url{https://csrc.nist.gov/publications/detail/fips/198/1/final}, 2008.

\bibitem{nist-sha}
National~Institute of~Standards and Technology.
\newblock Fips pub 180-4: Secure hash standard (shs).
\newblock \url{https://nvlpubs.nist.gov/nistpubs/FIPS/NIST.FIPS.180-4.pdf}, 2015.

\bibitem{abadi2016deep}
Martin Abadi, Andy Chu, Ian Goodfellow, H~Brendan McMahan, Ilya Mironov, Kunal Talwar, and Li~Zhang.
\newblock Deep learning with differential privacy.
\newblock In {\em Proceedings of the 2016 ACM SIGSAC Conference on Computer and Communications Security (CCS)}, pages 308--318, 2016.

\bibitem{uchida2017embedding}
Yusuke Uchida, Yuki Nagai, Shigeyuki Sakazawa, and Shin'ichi Satoh.
\newblock Embedding watermarks into deep neural networks.
\newblock In {\em Proceedings of the ACM International Conference on Multimedia Retrieval (ICMR)}, pages 269--277, 2017.

\bibitem{strogatz2024nonlinear}
Steven~H Strogatz.
\newblock {\em Nonlinear dynamics and chaos: with applications to physics, biology, chemistry, and engineering}.
\newblock Chapman and Hall/CRC, 2024.

\bibitem{ott2002chaos}
Edward Ott.
\newblock {\em Chaos in dynamical systems}.
\newblock Cambridge university press, 2002.

\end{thebibliography}

\end{document}